\documentclass{article}
\usepackage[round]{natbib}

\usepackage{amsthm}
\usepackage{color}
\usepackage{xcolor}
\usepackage{bm}
\usepackage{stmaryrd}
\usepackage{url}
\usepackage{amsmath}
\usepackage{listings}
\usepackage{graphicx}
\usepackage{multirow}
\usepackage{amssymb}
\usepackage{amsfonts}
\usepackage{authblk}
\usepackage[colorlinks=true, linkcolor=blue, citecolor=blue, urlcolor=blue]{hyperref}

\definecolor{codegreen}{rgb}{0,0.6,0}
\definecolor{codegray}{rgb}{0.5,0.5,0.5}
\definecolor{codepurple}{rgb}{0.58,0,0.82}
\definecolor{backcolour}{rgb}{0.95,0.95,0.92}
\lstdefinestyle{mystyle}{
    backgroundcolor=\color{backcolour},   
    commentstyle=\color{codegreen},
    keywordstyle=\color{magenta},
    numberstyle=\tiny\color{codegray},
    stringstyle=\color{codepurple},
    basicstyle=\ttfamily\footnotesize,
    breakatwhitespace=false,         
    breaklines=true,                 
    captionpos=b,                    
    keepspaces=true,                 
    numbers=left,                    
    numbersep=5pt,                  
    showspaces=false,                
    showstringspaces=false,
    showtabs=false,                  
    tabsize=2
}
\newcommand\R{{\bf R}}

\newcommand\E{{\bf E}}

\newtheorem{remark}{Remark}
\newtheorem{lemma}{Lemma}

\title{Model-Free Deep Hedging with Transaction Costs and Light Data Requirements}
\date{\today}
\author[ ]{Pierre Brugi\`ere \thanks{brugiere@ceremade.dauphine.fr, ORCID: \texttt{0000-0002-8716-9145}}}
\author[ ]{Gabriel Turinici  \thanks{turinici@ceremade.dauphine.fr, ORCID: \texttt{0000-0003-2713-006X}}}
\affil[ ]{CEREMADE, Universit\'e Paris Dauphine-PSL, 
 Place du Mar\'echal de Lattre de Tassigny, 75116 Paris FRANCE}

\begin{document}
	\maketitle

\begin{abstract}
Option pricing theory, such as the \citet{BlackScholes} model, provides an explicit solution to construct a strategy that perfectly hedges an option in a continuous-time setting. In practice, however, trading occurs in discrete time and often involves transaction costs, making the direct application of continuous-time solutions potentially suboptimal. Previous studies, such as those by \citet{Buehler1}, \citet{Buehler2} and \citet{Cao}, have shown that deep learning or reinforcement learning can  be used to derive better hedging strategies than those based on continuous-time models. However, these approaches typically rely on a large number of trajectories (of the order of $10^5$ or $10^6$) to train the model.
In this work, we show that using as few as 256 trajectories is sufficient to train a neural network that significantly outperforms, in the Geometric Brownian Motion framework, both the classical Black \& Scholes formula and the Leland model, which is arguably one of the most effective explicit alternatives for incorporating transaction costs. The ability to train neural networks with such a small number of trajectories suggests the potential for more practical and simple implementation on real-time financial series.
\end{abstract}

\section{Introduction}

In practice, continuously trading in response to all infinitesimal variations in an asset's price is not feasible. Therefore, perfect replication of an option as described in continuous-time models is not possible. 

Because trading occurs discretely, even if an asset follows a Geometric Brownian Motion (GBM)  with known parameters, as in the Black \& Scholes model, no real-world discrete hedging strategy allows for the perfect replication (almost surely on any path) of an option's payout.

In the first section, we discuss how discretization affects the accuracy of the hedging for  Black \& Scholes and other models. We remind the reader that the impossibility of perfect hedging is linked not only to the residual delta existing along the paths but also to the discretization of the gamma. Indeed, the gamma weighted discretized quadratic variation observed along a trajectory (the "gamma cash" generated by the option) can differ from its expected continuous value, just as the realized volatility measured from discrete points along a trajectory may differ from the volatility parameter of the GBM process that generated it.

Typically, the criterion used to define the best hedging strategy is based on the present value $Z_T$ of the final wealth achieved by the trader hedging the liability of having sold an option:
\begin{align}
  Z_T = \sum\limits_{i=0}^{n}\frac{C_i}{\beta_i},  
\end{align}
where $C_i$ is the cash flow for the trader a time $i$ and $\beta_i$ is the cash amount at time $i$ generated by  1 USD invested at the instantaneous risk free-rate between time $0$ and $i$.

 It can be formulated in three different ways: by maximizing a concave utility function, by minimizing a convex loss function, or by using a concave risk measure. In most cases, each formulation can be easily translated into one of the others, as explained by \citet{Follmer}, \citet{Xu} or \citet{Buehler1}.

In our approach, we choose to minimize the standard deviation of discounted terminal wealth $Z_T$, which is also \citet{KolmRitter} choice to define the optimal hedge, and we train a deep neural network to optimize it. This neural network is trained on a set of trajectories and tested on a different set.

Our approach differs from previous ones in the following aspects:
\begin{itemize}
\item we do not use paths simulated under a risk-neutral measure to calibrate our neural network;
\item we do not write our model using a  risk-free rate of zero;
\item we show that a lightweight model trained on no more than 256 trajectories is sufficient to achieve good performance (for a simple stochastic process). Furthermore, in our experiments, we find that we can calibrate our neural network with only 256 overlapping trajectories from 285 observation points.
\end{itemize}
Due to the small number of data points required to calibrate our model, we believe that these results pave the way for the use of deep hedging in certain real market situations, without the systematic need to enrich market data with synthetic observations generated by parametric models or model-free methods such as Generative Adversarial Networks \citep{Hirano} or Variational Autoencoder \citep{PBGT}. In fact, data augmentation using intraday data alone may be sufficient in some cases, as suggested by \citet{Mikkila}.

\section{Simulations Used}

The simulations used to train and test our model are generated under a real-world probability $\mathbb{P}$, which is not a risk-neutral probability $\mathbb{Q}$ associated with the risk-free rate used. For example, in the case of a GBM, we use a trend $\mu$ for the assets that differs from the risk-free rate $r$. Note that in continuous-time models, such as Black \& Scholes, the realized return of a trajectory is irrelevant to achieve perfect hedging; only the realized volatility matters.

Our optimization criterion is based on the discounted value $Z_T$ of the trader's final wealth at the expiration date of the option.

Concerning the number of paths required to calibrate a neural network model, we make the following remark. When analyzing the hedging of a call option strike $K$, if all paths in the training set lead to the option expiring worthless (i.e., the final asset price is below $K$ in every case), then the problem of optimizing $Z_T$ translates into finding an optimal trading strategy on the stock alone. 
In this case, the option liability $L=(X_T - K)^+$ is zero for all paths and thus does not play a role in the computation of $Z_T$, nor in the optimization process of the neural network.

Under such conditions, the neural network focuses solely on optimizing a stock trading strategy, without addressing any option-related objective. 

Since our optimization criterion is based on the standard deviation of $Z_T$, our result in this scenario would be not to take any position in the stock. However, other criteria, such as those based on mean-variance optimization or an entropic risk measure, could yield different results.

Because of this characteristic of the standard deviation criterion, we do not need to be overly concerned about having only a limited number of in-the-money paths or about the risk of the hedging strategy being "polluted" by a stock trading strategy exploiting statistical arbitrage.

Note that another way to determine the hedging strategy for an option with strike $K$, when the training set contains only a limited number of non-zero payouts, is to include options with different strikes $K_1, K_2, \dots, K_n$ that produce more nonzero payouts.

\section{States Variables and Hedging Strategies}
To better understand the challenges faced by the deep learning model within the discrete framework, as well as the problems of model or parameter mis-specification, we first discuss the replication method for continuous-time models in the absence of transaction costs. We consider two classic cases in which a set of state variables, evolving according to a Markovian process, fully characterizes the market state. In the first case, the asset follows a geometric Brownian motion (GBM) process, and the instantaneous interest rate $r_s$ is stochastic. In the second case, widely studied in the literature (see, e.g., \citet{Buehler1} and \citet{Cao}, etc.), the asset follows a Heston model. The approach of expressing the hedge in terms of state variables can be extended to other processes as well. \\

We note  $\beta_t$  be the stochastic discount factor
\begin{align}
\beta_t = exp(- \int_0^t  r_s ds),    
\end{align}
$B_t$  the zero coupon bond which pays $1$ at time $T$ and $C_t$ the price at time $t$ of a European option of maturity $T$. We assume that we can find a risk neutral probability $\mathbb{Q} $ under which:
\begin{align}
&dX_t = r_t X_t\, dt + \sigma (t,X_t)\, dW_t^\mathbb{Q}
\label{eq:risk_neutral_X}
\end{align}
where $(W_t^\mathbb{Q})_{t \geq 0}$ 
is a Brownian motion under $\mathbb{Q}$. In the absence of arbitrage opportunities,   
$  (\beta_t B_t)_{t\geq 0} $ 
and  $(\beta_t C_t)_{t\geq 0} $ are martingales under $\mathbb{Q}$.

We assume that we can write $B_t$ as a function of $(t,r_t)$ and $C_t$ as a function of $(t,X_t,r_t)$ and that these functions are differentiable for $t<T$. Note that this property is valid for most common models for $(r_s)_{s \geq 0}$

With these assumptions, it is easy to show that $C_t$ can be perfectly hedged by continuously trading $X_t$ and $B_t$ and that the replicating strategy is derived in a straightforward way from the hedging strategy. If we note $C_t=C(t,X_t,r_t)$  and $B_t=B(t,X_t)$ and $\E_t(.)$ the conditional expectation under $\mathbb{Q}$, knowing all the state variables up to times $t$, we get the following:
\begin{equation}
\left\{ 
\begin{array}{ll} 
dC_t=  r_tC_t dt + \frac{\partial C }{\partial x} (dX_t - \E_t(dX_t) )   +  \frac{\partial C }{\partial r} (dr_t - \E_t(dr_t) )  \\[10pt]
dB_t=  r_t B_t dt +  \frac{\partial B }{\partial r} (dr_t - \E_t(dr_t) )  
\end{array}\right.
\end{equation}

which leads to 
\begin{align}
dC_t- r_tC_t dt  =   \frac{\partial C }{\partial x} (dX_t - r_tX_tdt )   +  \frac{\partial C }{\partial r} \big( \frac{\partial B }{\partial r} \Big)^{-1} (dB_t - r_tB_t dt).
\label{eq:dcrcdt}
\end{align}

This implies that a short position in $C_t$ can be hedged by holding $\frac{\partial C }{\partial x} $ assets and $\frac{\partial C }{\partial r} \big( \frac{\partial B }{\partial r} \Big)^{-1}$ bonds and by financing (resp placing) any cash deficit (resp surplus) at the instantaneous risk-free rate $r_t$. 

To translate this hedging relationship into a replicating relationship, we multiply both terms in Equation \eqref{eq:dcrcdt} by $\beta_t$ 
and use that $d\beta_tC_t  = \beta_t( dC_t- r_tC_t dt) $
to obtain: 
\begin{align}
  d\beta_tC_t  =   \beta_t \frac{\partial C }{\partial x} (dX_t - r_tX_tdt )   +  \frac{\partial C }{\partial r} \big( \frac{\partial B }{\partial r} \Big)^{-1} \beta_t (dB_t - r_tB_t dt),   
\end{align}
and after integration between $0$ and $T$ and dividing by $\beta_T$:
\begin{align}
C_T  = \frac{ C_0 }{\beta_T}
+ \int_{0}^T \frac{\beta_t}{\beta_T} \frac{\partial C }{\partial x} (dX_t-rX_tdt) + \int_{0}^T \frac{\beta_t}{\beta_T} \frac{\partial C }{\partial r} \left( \frac{\partial B }{\partial r} \right)^{-1} (dB_t-rB_t dt).
\label{eq:decomp}
\end{align}
So, to replicate the option the positions in $X_t$ and $B_t$ are the same as for hedging, and the cash management strategy consists in investing or borrowing any cash surplus or deficit at the instantaneous rate $r_s$. Note that the replication equality holds for all paths almost surely and for all probabilities equivalent to the risk-neutral probability $\mathbb{Q}$ and thus under the real world probability $\mathbb{P}$ as well.

\vspace{0.5 cm}
For a \citet{Heston} model, 
Equation~\eqref{eq:risk_neutral_X} is not true any more; instead the two state variables are the asset price $X_t$ and the stochastic volatility $V_t$ which follow Equation (\ref{eq:Heston}) under the risk-neutral probability $\mathbb{Q} $:
\begin{equation} \label{eq:Heston}
\left\{ 
\begin{array}{ll} 
dX_t=r X_t dt + \sqrt{V_t} X_t dW^1_t\\
dV_t= \alpha(b-V_t)dt + \sigma \sqrt{V_t}  dW^2_t \\
W^1_t, W^2_t \mbox{ Brownian motions under } \mathbb{Q}
\end{array}\right.
\end{equation}

The reasoning for replication is similar; the possibility to trade $X_t$ and an asset $S_t=S(t,V_t)$ enables a perfect replication in continuous time. The asset $S_t$ could for example be a variance swap starting at time zero and whose final payout at  time $T$ is,
$$S_T= \int_0^T e^{r(T-s)}V_sds.$$ 

In the absence of arbitrage opportunities, the valuation at times $t$ of $S_t$ is $S_t=\E_t \big( e^{-r(T-t)}  S_T  \big)$. In this case, the replicating strategy can be written as 

\begin{equation}
\begin{aligned}
C_T  &= e^{rT}C_0  + \int_{0}^T e^{r(T-t)} \frac{\partial C }{\partial x} (dX_t -rX_t dt)  \\
     &\quad + \int_{0}^T  e^{r(T-t)}  \frac{\partial C }{\partial V} \left( \frac{\partial S }{\partial V} \right)^{-1} (dB_t -rB_t dt)
\end{aligned}
\end{equation}

This is the same as in \citet{Buehler1} for $r=0$. Note that we will not test the hedging accuracy of our neural network in a Heston model, as we would need to assume that we have access to a perfectly priced and tradable instrument such as $S_t$, which may not be the case.

We will first analyze the effect of discretization from a theoretical point of view for Black \& Scholes, both without transaction costs and with transaction costs, as in Leland's work \citep{Leland}, before implementing our neural network and discussing the results obtained. We will show that the results are consistent with the theory, and the possibility to calibrate the hedging on a very limited number of trajectories seems promising for implementation in real market conditions.

\section{ Realized Volatility and Hedging Errors}

In discrete time, the realized volatility computed from a finite number of observations may differ from the volatility used in the simulation of the process. Before analyzing the impact of discretization on hedging performance, we first examine the effect of volatility misspecification in a continuous-time setting. This analysis is carried out in the simplest setting of the Black \& Scholes model.

We assume that the risk-free interest rate is $r$ and that under the real world probability $\mathbb{P}$ the asset price follows the stochastic differential equation:
\begin{align}
dX_t &=  \mu X_tdt + \sigma X_t dW_T^\mathbb{P},    
\end{align}
where  $(W_t^\mathbb{P})_{t \geq 0}$ is a Brownian motion under $\mathbb{P}$.

Let $C_{\nu}(t,x)$ denote the Black \& Scholes price at time $t$ of a European call option that pays $C_{\nu}(T,x) = (x -K)^+$ at maturity,  where the pricing uses the implied volatility $\nu$
and the interest rate  $r$. We have 
$$\frac{\partial C_{\nu}}{\partial t} + \frac{\partial C_{\nu}}{\partial x}rx + \frac{1}{2}\nu^2 x^2   \frac{\partial^2 C_{\nu}}{\partial^2 x} =rC_{\nu}.$$

Leaving $C^{\nu}_t= C_{\nu}(t,X_t)$ and $Y_t^{\nu}=e^{-rt}C_t^{\nu}$, we have the following.

\begin{equation}
\label{eq:C}
\begin{aligned} 
dY_t^{\nu}= -re^{-rt}C_t^{\nu} dt + e^{-rt}\frac{\partial  C_{\nu}}{\partial  t}dt  + e^{-rt}\frac{\partial  C_{\nu}}{\partial  x}dX_t +\frac{1}{2} e^{-rt} \sigma^2 x^2  \frac{\partial^2 C_{\nu}}{\partial^2 x}dt \\
=  e^{-rt}\frac{\partial  C_{\nu}}{\partial  x}(dX_t -rX_tdt) +\frac{1}{2}  e^{-rt} ( \sigma^2 - \nu^2  )x^2  \frac{\partial^2 C_{\nu}}{\partial^2 x}dt.
\end{aligned}
\end{equation}

So, by integration between $0$ and $T$,
\begin{equation} \label{eq:hedge}
\begin{aligned}   e^{-rT}C_T - C_0^{\nu} =  \int_0^T e^{-rt}  \frac{\partial C_{\nu}}{\partial x}(X_s,s)(dX_s-rX_sdt) \\
   + \frac{1}{2} \int_0^T e^{-rt} ( \sigma^2 - \nu^2  )X_t^2\frac{\partial^2 C_{\nu}}{\partial^2 x}(X_s,s)dt.
\end{aligned} 
\end{equation}

 So, a trader hedging the liability $C_T$  using for Black \& Scholes  the implied volatility $\nu$ - with the hedge being financed at the rate $r$ and any gain or loss realized  being invested (resp financed) at the rate $r$ - would have for $Z_T$:
 \begin{equation}
 \label{eq:sigma}
\begin{aligned} 
Z_T =    \int_0^T e^{-rt}  \frac{\partial C_{\nu}}{\partial x}(X_s,s)(dX_s-rX_sdt)  - e^{-rT}C_T \\
= \frac{1}{2}  (  \nu^2  - \sigma^2 ) \int_0^T e^{-rt} \frac{\partial^2 C_{\nu}}{\partial^2 x}(X_s,s)X_t^2 dt -C_0^{\nu}.    
\end{aligned} 
 \end{equation}
 
From this we get the known results that:
\begin{itemize}
\item  if the implied volatility $\nu$ is equal to the asset's true volatility $ \sigma$, the $\nu$-hedge is perfect. In this case, the present value of the cost of hedging the position is deterministic and equal to $C_0^{\nu}$, which is the option premium at inception.
\item if the  implied volatility $\nu$ is superior to the asset true volatility  $ \sigma$ , then since the  gamma $\frac{\partial^2 C_{\nu}}{\partial^2 x} $  is always positive,  
a trader who received the premium  $C_0^{\nu}$  at inception and who did  a $\nu$-hedge is guaranteed a gain at maturity. However, the exact amount of the gain depends on the gamma encountered throughout the hedging period.
 \end{itemize}

Note that the result in Equation (\ref{eq:sigma}) does not depend on the actual drift $\mu$ of the asset price $ X_t$ under the real-world probability $\mathbb{P}$  -  as expected when pricing in a risk-neutral probability framework. From a simulation perspective, this implies that we can expect to find an effective hedging strategy even if the asset paths are not generated under the risk-neutral probability.

We now examine the effect of discretization on hedging, both with and without transaction costs, to understand how this will affect our deep hedging strategy.

\subsection{Hedging Discretely with No Transaction Costs}

First, we examine the impact of discrete hedging (and realized discrete volatility) on hedge accuracy. If, at the discrete hedging points $0=t_0<t_1, \cdots <t_n=T$, the delta used is the Black \& Scholes delta for an implied volatility $\nu$ (which we do not necessarily assume equal to $\sigma$), then the present value $Z_T$ of hedging the liability, while financing (or investing) the cash deficit (or surplus) at rate $r$, is:

\begin{equation}
\begin{aligned}
\label{demo:appendix}
Z_T &= \sum\limits_{i=0}^{n-1}     e^{-rt_{i+1}}      \frac{\partial C_{\nu}}{\partial x}(X_{t_i},t_i) 
  \Big[X_{t_{i+1} } - e^{r( t_{i+1}  -t_i   )}X_{t_i}  \Big] - e^{-rT}C_T. \\
\end{aligned}
\end{equation}

If we assume that the hedging takes place at regular time intervals, we have $t_{i+1}-t_i = \Delta t = \frac{T}{n}$ and we can write as shown in Appendix \ref{appendix:proof}:

\begin{equation} \label{eq:approx1}
 Z_T=    \frac{1}{2}  \sum\limits_{i=1}^n e^{-rt_{i+1}} \frac{\partial^2 C_{\nu}}{\partial^2 x} (X_{t_i},t_i) X_{t_i}^2  \Big[  \nu^2 (t_{i+1} -t_i  ) -\big(\frac{X_{t_{i+1}} - X_{t_i}}{X_{t_i}}\big)^2  \Big]  - C_0^{\nu} +O(\sqrt{\Delta t}),
 \end{equation}
 
 which is (when $\nu = \sigma $) the known Leland's formula \citep{Leland}.  When $\Delta t $ tends to zero, we obtain Equation (\ref{eq:sigma}), which we derived in the continuous case.
 Thus, the uncertainty associated with discrete hedging is, at the first order, linked to the difference between:
 
- the gamma cost incurred by the trader:

 $$  \frac{1}{2}  \sum\limits_{i=1}^n e^{-r(T-t_i)} \frac{\partial^2 C_{\sigma}}{\partial^2 x}  X_{t_i}^2   \big(\frac{X_{t_{i+1}} - X_{t_i}}{X_{t_i}}\big)^2   $$

- and its cumulative discounted expected value  
$$
  \frac{1}{2}  \sum\limits_{i=1}^n e^{-r(T-t_i)} \frac{\partial^2 C_{\sigma}}{\partial^2 x} (X_{t_i},t_i) X_{t_i}^2    \sigma^2 (t_{i+1} -t_i  ). $$

The variability of $Z_T$  - even when $\nu=\sigma$ -  arises from the fact that the realized (squared) volatility measured from discrete time points along a trajectory (and gamma-weighted) is merely an estimator of $\sigma$ and thus a random variable. Only in the case of continuous-time hedging, where the realized volatility converges almost surely to $\sigma$, does $Z_T$ correspond to a perfect hedge.

In our training set, which included 30 or 90 observations per trajectory, the estimator of $\sigma^2$ can differ significantly from its limit value. For this simple reason, perfect hedging is not achieved. Indeed, in our sample of 256 trajectories simulated with $\sigma=20\%$, the estimated values of $\sigma$ computed with 30 observation points vary between $12.51\%$ and $26.89\%$.

\begin{remark}
The time value $V_t$ represents the difference between the price of the option and its intrinsic value $(X_t-K)^+ $, so:
$$ V_t = C_t - (X_t-K)^+$$ 
When selling an option, a trader incurs some gamma costs, but benefits from the reduction in the time value of the option $V_t$. If the option moves deeply out of the money, its gamma approaches zero, which may seem favorable to the trader as gamma costs diminish.  However, this is exactly offset by the fact that the time value approaches zero as well, meaning that the trader will no longer benefit from its time decay. When the option remains near the money, both gamma and time value are high, and once again the decay of the time value tends to compensate for the gamma costs. As the option approaches maturity, if it remains at-the-money (ATM), gamma reaches its maximum. In this case, managing delta becomes increasingly difficult, as seen in the simulation where the hedging frequency is fixed and proves insufficient when the option is close to the money at expiry.

\end{remark}

\subsection{Hedging Discretely with Transactions Costs}

It is in this setting that our deep learning model performs significantly better than the Black \& Scholes  delta hedge.
Define $\Delta t = T/n$ and $t_i = i \cdot \Delta t$. 
Our hedging strategy $H_{t+\Delta t}$ at time $t+ \Delta t $ which is our holding in the asset is modeled as a function of $t+\Delta t$, $H_{t}$ (the pre-existing asset position) and $X_{t+\Delta t}$.  This time, we compare our model to the Black \& Scholes model and to the Black \& Scholes model adjusted according to Leland's formula \citep{Leland}.

We assume that the trading costs are proportional to the value traded.
We examine the transaction costs incurred by a trader who hedges discretly using the Black \& Scholes model with implied volatility $\nu$, which may differ from the volatility of the asset price $\sigma$. Trading costs are assumed to apply, neither at inception, when the trader exchanges the delta with the option buyer, nor at maturity for the stocks which are delivered against the strike, since the option is assumed to be physically settled.  Thus, the transaction costs of $\alpha$ per dollar traded arises only when the deltas are readjusted at times  $t_{i+1}$ and are equal to:  
\begin{equation} 
\label{eq:t_costs}
\alpha \Big |    \frac{\partial C_{\nu}}{\partial x}(X_{t_{i+1}}, t_{i+1}) - \frac{\partial C_{\nu}}{\partial x}(X_{t_{i}},t_i)  \Big |  X_{t_{i+1}} ,
\end{equation}

which can be expressed as, 
$$  \alpha  \frac{\partial^2 C_{\nu}}{\partial^2 x}(t_i,X_{t_{i}}) | X_{t_{i+1}} - X_{t_i}| X_{t_i} + O(\Delta t).$$

Following  \citet{Leland} if we assume that $\nu$ depends on $\Delta t$ as in,   \begin{equation}
\label{modified_vol}
\nu^2= \sigma^2 +\frac{\alpha}{\sqrt{\Delta t}}  \mbox{ with }  \alpha >0,
\end{equation}
 we get,
 $$ \frac{\partial^2 C_{\nu}}{\partial^2 x}(t_i,X_{t_{i}}) = O(\sqrt{\Delta t)}$$
and the transactions costs in Equation (\ref{eq:t_costs}) can be expressed as,
$$  \alpha  \frac{\partial^2 C_{\nu}}{\partial^2 x}(t_i,X_{t_{i}}) | X_{t_{i+1}} - X_{t_i}| X_{t_i} + O(\Delta t^{\frac{3}{2}}).$$
So, in this case, the cumulated discounted transaction costs are: 

\begin{equation}
\label{eq:ourapprox} \alpha  \sum\limits_{i=0}^{n-1} e^{-rt_{i+1}} \frac{\partial^2 C_{\nu}}{\partial^2 x}(t_i,X_{t_{i}}) | X_{t_{i+1}} - X_{t_i}| X_{t_i} +  O(\sqrt{\Delta t}). 
\end{equation}

Note that, for any fixed $\nu$, the gamma sum 
in Equation (\ref{eq:ourapprox}) tends to infinity 
when $\Delta t$ tends to zero; see \citet{Soner}. 
In fact, for any fixed $\nu$, hedging using Black \& Scholes deltas, in the presence of proportional transaction costs, leads to infinite transaction costs, when $\Delta t$ tends to zero. 
We present in Appendix \ref{appendix:proof2} a short proof of this matter. 

With Leland's volatility adjustment, the volatility increase reduces the gamma, which keeps the transaction costs under control in Equation (\ref{eq:ourapprox}).
Assuming that $\nu$ satisfies Equation \eqref{modified_vol} we get from \citet{Leland} that

$$ \sum\limits_{i=0}^{n-1} e^{-rt_{i+1}} \frac{\partial^2 C_{\nu}}{\partial^2 x}(t_i,X_{t_{i}}) | X_{t_{i+1}} - X_{t_i}| X_{t_i} $$
$$ = \alpha  \sum\limits_{i=0}^{n-1} e^{-rt_{i+1}} \  \frac{\partial^2 C}{\partial^2 x}(t_i,X_{t_{i}}) \sigma X^2_{t_i}\sqrt{ \frac{2\Delta T}{ \pi}} + O(\sqrt{\Delta t}). $$ 

Taking these transaction costs into account in Equation (\ref{eq:approx1}), we obtain as the total hedging costs for the trader:
\begin{equation} 
\begin{aligned}
 Z_T=    \frac{1}{2}  \sum\limits_{i=1}^n e^{-rt_{i+1}} \frac{\partial^2 C_{\nu}}{\partial^2 x} (X_{t_i},t_i) X_{t_i}^2  \Big[  \nu^2 (t_{i+1} -t_i  ) -\big(\frac{X_{t_{i+1}} - X_{t_i}}{X_{t_i}}\big)^2  - \alpha  \sigma  \sqrt{ \frac{2\Delta t}{\pi}} \Big] 
 \\ - C_0^{\nu} +O(\sqrt{\Delta t}).
\end{aligned}
\end{equation}
 
So, if $\nu^{*2 }= \sigma^2 +\alpha  \sigma  \sqrt{ \frac{2}{\pi \Delta t}} = \sigma^2 +\alpha  \sigma  \sqrt{ \frac{2n}{\pi T}}$ in Equation (\ref{modified_vol}) we get:
\begin{equation} 
 Z_T=    \frac{1}{2}  \sum\limits_{i=1}^n e^{-rt_{i+1}} \frac{\partial^2 C_{\nu^*}}{\partial^2 x} (X_{t_i},t_i) X_{t_i}^2  \Big[  \sigma^2 (t_{i+1} -t_i  ) -\big(\frac{X_{t_{i+1}} - X_{t_i}}{X_{t_i}}\big)^2   \Big]  - C_0^{\nu^*} 
+O(\sqrt{\Delta t}).
 \end{equation}

Therefore, Leland's adjusted volatility $\nu^*$ appears to be a good candidate as the implied volatility to hedge the option and to reduce the uncertainty associated with the gamma costs. The smaller $\Delta t$, the higher $\nu^*$ is, and the smaller the gammas.

\begin{remark}
In Leland's analysis, the delta does not depend on the existing hedging position, whereas in our neural network strategy it will, as discussed in the next section.
\end{remark}

\section{Simulations}
\label{sec:simul}
We build observation points for trajectories following a GBM process by generating log-normal variables that represent the evolution of asset prices between successive observation points. The parameters for the simulated trajectories are as follows:
\begin{lstlisting}
# Parameters for the trajectories
S0 = 1                 # Initial stock price
mu = 0.05              # Drift for the asset price
sigma = 0.2            # Volatility of the asset price
T= 0.25                # Time to maturity, as a fraction of a year
steps = 30             # or 90, Number of time intervals 
num_paths = 256        # Number of simulated asset paths
seed_value = 42        # Seed value for reproductibility
\end{lstlisting}

The realized volatilities $\hat{\sigma}_i$ computed on 30 hedging points for the 256 trajectories range from $12.51\%$  to $26.89\%$  with  a mean of $19.66\%$ and a standard deviation of $2.53\%$ (for 90 hedging points, the minimum value is 16.00\%, the maximum value 25.18\%, the average value 19.87\% and the standard deviation is $1.53\%$). Based on these results and Equation (\ref{eq:approx}), it is expected that the discrete hedging will be far from perfect, since the realized volatilities computed at the hedging points differ from one path to another and can deviate substantially from the $20\%$ average volatility used to generate trajectories and compute the Black \& Scholes deltas. As a result, any model that applies the same delta function for all paths will suffer from this limitation.

\begin{figure}[h]
    \centering
    \includegraphics[width=1.0\textwidth]{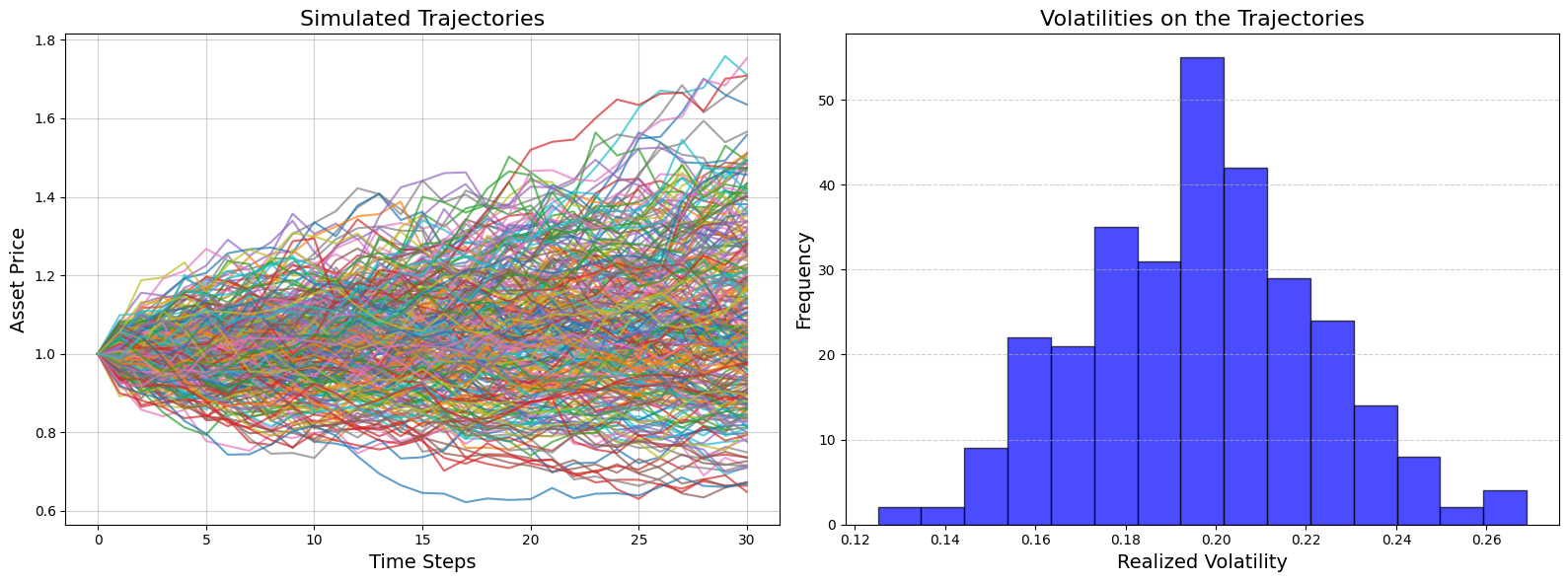}
    \caption{Simulations and realized volatilities for 30 time steps, as discussed in section \ref{sec:simul}.}
    \label{fig:mon_image1}
\end{figure}

\subsection{Neural Network model used}
We use a neural network to optimize hedging under proportional transactions cost $\alpha$. The case of zero transaction costs is a special case and does not require a distinct neural network.
With transaction costs, the situation is as follows: If $M_t$ denotes the cash position at time $t$ and $H_{t-\Delta t}$ the number of assets held at time $t-\Delta t$, then the valuation at time $t$, before hedge adjustment, is:
$V_t = M_t+H_{t-\Delta t} X_t$. Once the hedge is adjusted to $H_t$, the next valuation of the position at time $t+\Delta t$ will be:
$$
\begin{aligned}
V_{t+\Delta t} = e^{r\Delta t}M_t -   (H_t - H_{t-\Delta t}) X_t e^{r\Delta t} -\alpha .| H_t - H_{t-\Delta t}|.X_t e^{r\Delta t} +      H_{t} X_{t+\Delta t} \\
= e^{r\Delta t} (M_t  +H_{t-\Delta t}X_t)  -\alpha .| H_t - H_{t-\Delta t}|.X_t e^{r\Delta t}  + H_{t}(X_{t+\Delta t} - e^{r\Delta t}X_t) \\
=  e^{r\Delta t} V_t   - \alpha .| H_t - H_{t-\Delta t}|.X_t e^{r\Delta t}  + H_{t}(X_{t+\Delta t} - e^{r\Delta t}X_t)
\end{aligned}
$$ 
So, when there are no transaction costs ($\alpha=0$) $V_{t+\Delta t}-  e^{r\Delta t} V_t$ depends only on $H_t$. However, when transaction costs occur, it also depends on $H_{t-\Delta t}$. For this reason, the neural network that determines the number $H_t$ of shares to hold at time  $t$  is based on the values of $ (t,X_t,H_{t-\Delta t})$.
\vspace{0.5cm}

In practice, the neural network $\delta_M$ uses as state variables: 
\begin{itemize}
\item the moneyness of the option $\frac{K}{X_t}$  
\item the time left to maturity $T-t$
\item the pre-existing hedge position $H_{t-\Delta t}$ 
\end{itemize}

The neural network, written in Pytorch, combines "linear layers" and fully connected with "relu activations". We put no constraint on the output of the last layer, to be between zero and one.

\begin{lstlisting}
#######        Neural Network for model_delta   ###################                      
class DeltaNN(nn.Module): # input [batch_size, 3]
    def __init__(self):
        super(DeltaNN, self).__init__()
        self.fc1 = nn.Linear(3, 64)  
        self.relu = nn.ReLU()        
        self.fc2 = nn.Linear(64, 32) 
        self.fc3 = nn.Linear(32, 1)  
    def forward(self, x):
        x = self.fc1(x)
        x = self.relu(x)
        x = self.fc2(x)
        x = self.relu(x)
        x = self.fc3(x)
        return x          # output [batch_size,1] 
# Instantiate the model
model_delta = DeltaNN() 
\end{lstlisting}

For the transaction costs, we assume that they do not apply to the initial delta, as this is the case when the option buyer delivers the initial delta to the option seller, and that they do not apply at maturity to the shares that are delivered under physical exercise of the option. The present value $Z_T$ of the trader's  P\&L is calculated taking into account a fixed interest rate $r$ for financing the hedge. During the option life, when the delta increases the additional financing to buy more shares takes place at the same rate $r$ and when the delta decreases the value of the shares sold reduces the financed amount. The present value $Z_T$ of the cash flows generated by the trader hedging his option liability is as follows:

\begin{lstlisting}    
### PV hedging costs on trajectory_Y with hedging strategy delta
# as a % of initial notional p_0
# delta= [d_0,....d_29]  
# trajectory_Y=[p_0,...p_30]
# r: risk free rate
# K: moneyness at inception
def cost_Y_M (delta, trajectory_Y, r, K):
  n=len(trajectory_Y)-1        # n = 30                
  Y_t= trajectory_Y                               
  gamma=torch.exp(torch.tensor(-r * float(dt))) # discount factor
  cost_value = delta[0]        # initial delta cost
  for i in range(1, n):        # times [1,29]
  # intermediaries hedge adjustments
    cost_value = cost_value + (gamma ** i) * ((delta[i] - delta[i - 1]) + tc*torch.abs(delta[i]-delta[i - 1]))* Y_t[i]/Y_t[0]
  if Y_t[n]/Y_t[0]<K: # If the option is out of the money at maturity
    # liquidate the long position if any
    cost_value = cost_value -(gamma ** n) * torch.relu(delta[n-1])*(1-tc)* Y_t[n]/Y_t[0]
    # re-purchase the short position if any
    cost_value = cost_value +(gamma ** n) * torch.relu(-delta[n-1])*(1+tc)* Y_t[n]/Y_t[0]
  else: # If the option is in the money at maturity
    # receives the strike
    cost_value = cost_value - (gamma ** n) * K
    # purchase shares for delivery if holding less than 1
    cost_value = cost_value + (gamma ** n) *torch.relu(1-delta[n-1]) * Y_t[n]/Y_t[0]*(1+tc)
    # liquidate the shares if holding more than 1
    cost_value = cost_value - (gamma ** n) *torch.relu(delta[n-1]-1) * Y_t[n]/Y_t[0]*(1-tc)
  return cost_value
\end{lstlisting}

For optimization, the weights of the neural network are initialized with the standard "Kaiming Uniform" of Pytorch. Batches of 64 trajectories are used to optimize the parameters of the neural network. The loss to minimize is calculated for each batch as the standard deviation of $Z_T$. $Z_T$ is the Present Value of the P\&L at maturity of the trader who is hedging a short position in the option of final payout $(X_T-K)^+$. The optimizer is Adam with parameter $10^{-3}$.

\begin{lstlisting}                   
optimizer = optim.Adam(model_delta.parameters(), lr=0.001)
batch_size = 64
\end{lstlisting}

\begin{figure}[h]
    \centering
    \includegraphics[width=0.7\textwidth]{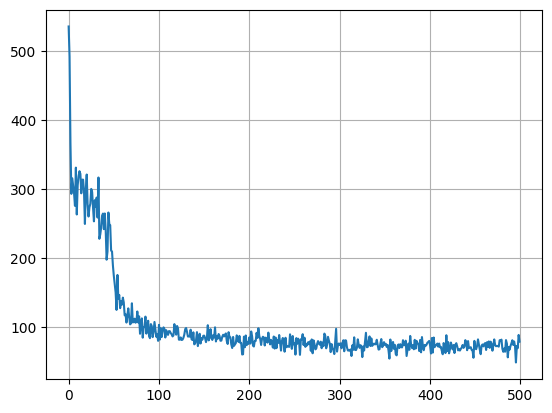}
    \caption{Calibration of Independent Sequences: standard deviations in basis points of the hedging costs $Z_T$ with 30 hedging points and $T=0.25$ for the 500 epochs, , as discussed in section \ref{sec:without}.}
    \label{fig:loss}
\end{figure}

\begin{figure}[h]
    \centering
    \includegraphics[width=1.0\textwidth]{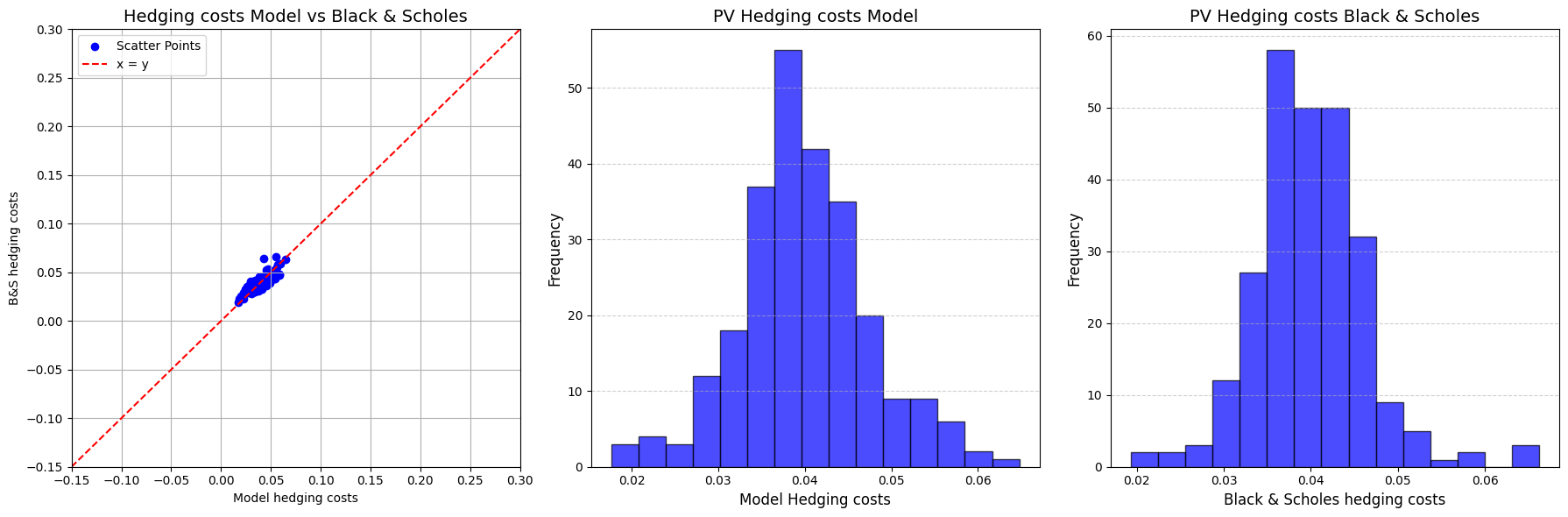}
    \caption{Test Set, as discussed in section \ref{sec:without}: hedging costs $Z_T$ with 30 hedging points and $T=0.25$ for the 256 trajectories considered.}
    \label{fig:histo}
\end{figure}

For each case studied, after 500 epochs, which takes a few minutes under Google Colab with no GPU usage, the neural network training is completed (see Figure \ref{fig:loss}).
The results obtained for the neural network are compared to those obtained with the standard Black \& Scholes hedge and with a Black \& Scholes hedge using an implied volatility given by the Leland's formula. We consider transaction cost rates of $0\% $, $0.2 \%$, $0.5\%$, $1\%$ and $2\%$. The test set is made of 256 trajectories independently generated of various lengths depending on the case considered. The results for the train set and test set are presented in Tables \ref{tab:1} and Tables \ref{tab:2}. In Figure \ref{fig:histo} we can see the distribution of the costs of hedging $Z_T$ for the neural network and Black \& Scholes when there are no transaction costs.

\subsection{Results Without Transaction Costs}
\label{sec:without}
\begin{figure}[h]
    \centering
    \includegraphics[width=1.0\textwidth]{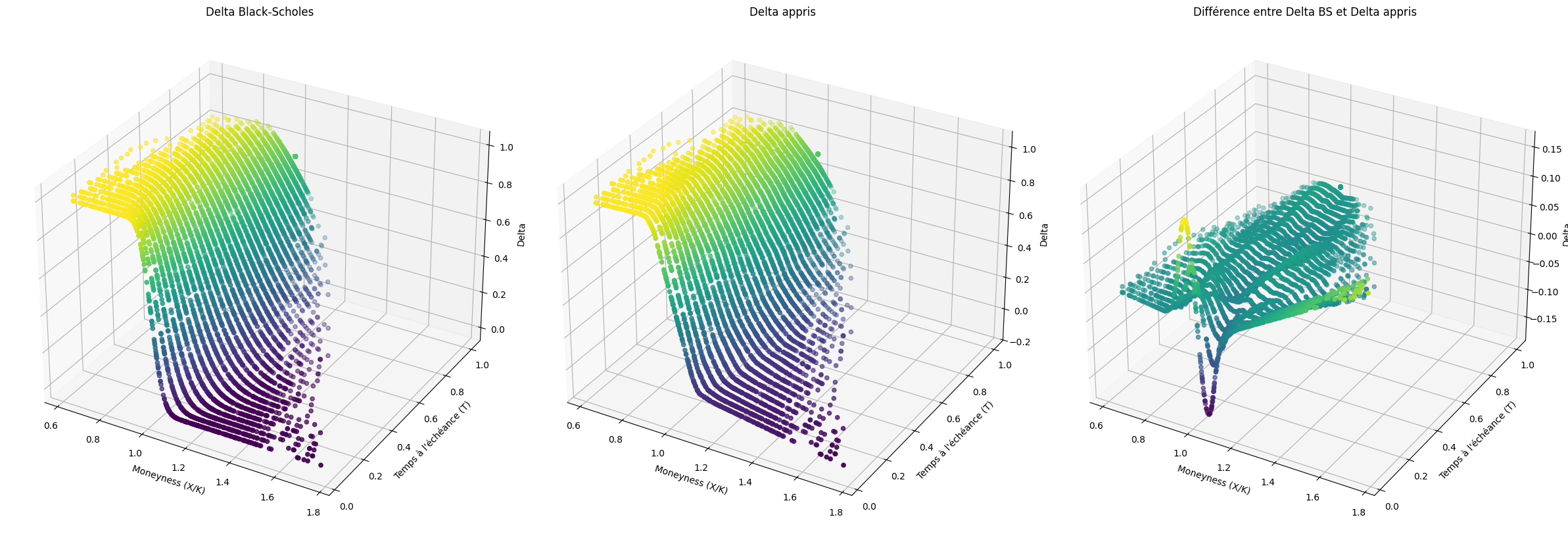}
    \caption{Realized Black \& Scholes deltas, Neural Network deltas, and differences, for all the hedging points of the 256 trajectories, as discussed in section \ref{sec:without}.}
    \label{fig:mon_image2}
\end{figure}

 When we compare in Figure \ref{fig:mon_image2} the delta surfaces produced by the Black \& Scholes model and the neural network at the points $(K_i,\tau_i, H_i)$ visited by the 256 paths in the training set, we observe that they are very similar. The only significant difference occurs for at-the-money options one time step before maturity, which is expected as the delta tends to be highly unstable in such cases.
As a result, the mean hedging values and the standard deviation of the hedging values can be considered to be nearly identical for Black \& Scholes and the neural network in these cases.

The differences in results between the training set and the test sets are similar for both the Black \& Scholes and the neural network. These discrepancies arise because, as each set contains only a limited number of trajectories, the statistics derived from them, such as the average and the standard deviation of realized volatilities, can differ slightly between the two sets. The key point to highlight is that, with our approach, the neural network does not require a large number of trajectories or observation points.

\begin{remark}
When learning to hedge a particular option, the neural network estimates values $C(K_i,\tau_i,H_i) $ that are also useful to hedge other options. Therefore, learning to hedge one option can help in hedging others. 
This observation is particularly important in cases where, due to the limited number of paths, a specific option never finishes in the money in the training set. In such scenarios, the neural network might incorrectly learn that always using a delta of zero is optimal, as this leads to a profit and loss (P\&L) standard deviation of zero - clearly not a desirable or realistic outcome.
To address this, one can train the network to hedge multiple option payoffs simultaneously by minimizing the sum of the standard deviations of the hedging costs $Z_T$ across $k$ different options, as expressed in Equation (\ref{eq:solve}):
 
\begin{equation} \label{eq:solve}
stdev (Cost_1)+ \cdots + stdev(Cost_k) \end{equation} 
In the cases presented here, this additional step was not necessary as we observed similar convergence results regardless of the additional options added to the one being hedged.
\end{remark}

When increasing the hedging frequency from 30 to 90 points, we observe a small change - approximately 5 basis points on average - in the average value of the hedging costs. This change is due to slight differences in the statistical properties and average realized volatilities of the trajectory sets. As expected, the standard deviation of the hedging cost decreases with increased hedging frequency, since realized volatilities are less dispersed and residual delta risk is reduced. Naturally, as the number of hedging points tends to infinity, the standard deviation of the hedging costs tends towards zero, and the average cost converges to the Black \& Scholes value.

\begin{remark}
Note that when analyzing the effect of discretization, it would likely be more relevant for practitioners to discretize not with respect to time but based on changes in the underlying asset - for example, implementing a hedge whenever the underlying moves by 1\%. This approach can be implemented without significant additional complexity, but is not addressed in this work.
\end{remark}

\begin{table}[h]
  \centering
  \caption{
  Independent Sequences, as discussed in section \ref{sec:Results}: \\ Call Strike 100\%, $T=0.25$,  freq = 30. 
  \\ Mean and Standard Deviation for $Z_T$ when $\sigma = 20\%$ and $r=0$.} 
  \vspace{5pt}  
  \label{tab:1}
  \begin{tabular}{|c|c|c|c|c|}
    \hline
    $\alpha$ & Costs for 256 Paths & Black \& Scholes & \textbf{Our Neural Network}     & Leland \\
        \hline
    \multicolumn{5}{|c|}{\textbf{Train Set}} \\
            \hline
    \multirow{2}{*}{0\% } & Mean & 4.04\% & 4.00\% & 4.04\% \\
                           & Standard Deviation & 0.64\% & 0.72\% & 0.64\%  \\
    \hline
    \multirow{2}{*}{0.2\% } & Mean & 4.40\% & 4.36\% & 4.39\% \\
                           & Standard Deviation & 0.68\% & 0.76\% & 0.67\%  \\
    \hline
        \multirow{2}{*}{0.5\% } & Mean & 4.95\% & 4.88\% & 4.91\% \\
                           & Standard Deviation & 0.79\% & 0.77\% & 0.72\%  \\
  \hline
          \multirow{2}{*}{1\% } & Mean & 5.86\% & \textbf{5.66\%} & 5.71\% \\
                           & Standard Deviation & 1.05\% & \textbf{0.80\%} & 0.82\%  \\
  \hline
            \multirow{2}{*}{2\% } & Mean & 7.68\% & \textbf{6.90\%} & 7.18\%\\
                           & Standard Deviation & 1.68\% & \textbf{0.87\%} & 1.03\% \\
  \hline
      \multicolumn{5}{|c|}{\textbf{Test Set}} \\
            \hline
      \multirow{2}{*}{0\%} & Mean & 3.99\% & 3.99\% & 3.99\% \\
                           & Standard Deviation & 0.67\% & 0.75\% & 0.67\%  \\
    \hline
    \multirow{2}{*}{0.2\%} & Mean & 4.35\% & 4.36\% & 4.34\% \\
                           & Standard Deviation & 0.67\% & 0.78\% & 0.66\%  \\
    \hline
        \multirow{2}{*}{0.5\%} & Mean & 4.89\% & 4.87\% & 4.84\% \\
                           & Standard Deviation & 0.76\% & 0.83\% & 0.70\%  \\
  \hline
          \multirow{2}{*}{1\%} & Mean & 5.79\% & 5.61\% & 5.64\% \\
                           & Standard Deviation & 0.98 \% & 0.87\% & 0.79\%  \\
  \hline
            \multirow{2}{*}{2\%} & Mean & 7.60\% & \textbf{6.83\%} & 7.09\% \\
                           & Standard Deviation & 1.55 \% & \textbf{0.97\%} & 0.98\%  \\
  \hline       
  \end{tabular}
\end{table}

\begin{table}[ht]
  \centering
  \caption{Independent Sequences, as discussed in section \ref{sec:Results}: \\ Call Strike 100\%, $T=0.25$,  freq = 90 
    \\ Mean and Standard Deviation for $Z_T$ when $\sigma = 20\%$ and $r=0$.} 
  \vspace{5pt}  
  \label{tab:2}
  \begin{tabular}{|c|c|c|c|c|}
    \hline
    $\alpha$ & Costs for 256 Paths & Black \& Scholes & \textbf{Our Neural Network}     & Leland \\
    \hline
    \multicolumn{5}{|c|}{\textbf{Train Set}} \\
            \hline
    \multirow{2}{*}{0\% } & Mean & 3.99\% & 3.80\% & 3.99\% \\
                           & Standard Deviation & 0.38\% & 0.45\% & 0.38\%  \\
    \hline
    \multirow{2}{*}{0.2\% } & Mean & 4.60\% & 4.54\% & 4.58\% \\
                           & Standard Deviation & 0.45\% & 0.47\% & 0.41\%  \\
    \hline
        \multirow{2}{*}{0.5\% } & Mean & 5.52\% & 5.27\% & 5.40\% \\
                           & Standard Deviation & 0.72\% & 0.50\% & 0.49\%  \\
  \hline
          \multirow{2}{*}{1\% } & Mean & 7.06\% & \textbf{6.27\%} & 6.64\% \\
                           & Standard Deviation & 1.27\% & \textbf{0.60\%} & 0.65\%  \\
  \hline
            \multirow{2}{*}{2\% } & Mean & 10.13\% & \textbf{8.04\%} & 8.79\% \\
                           & Standard Deviation & 2.45\% & \textbf{0.71\%} & 0.95\%  \\
  \hline
      \multicolumn{5}{|c|}{\textbf{Test Set}} \\
            \hline
      \multirow{2}{*}{0\%} & Mean & 3.96\% & 3.94\% & 3.96\% \\
                           & Standard Deviation & 0.37\% & 0.49\% & 0.37\%  \\
    \hline
    \multirow{2}{*}{0.2\%} & Mean & 4.59\% & 4.51\% & 4.56\% \\
                           & Standard Deviation & 0.42\% & 0.50\% & 0.38\%  \\
    \hline
        \multirow{2}{*}{0.5\%} & Mean & 5.54\% & 5.25\% & 5.39\% \\
                           & Standard Deviation & 0.66\% & 0.55\% & 0.45\% \\
  \hline
          \multirow{2}{*}{1\%} & Mean & 7.13\% & \textbf{6.26\%} & 6.64\% \\
                           & Standard Deviation & 1.19 \% & \textbf{0.60\%} & 0.60\%  \\
  \hline
            \multirow{2}{*}{2\%} & Mean & 10.13\% & \textbf{8.11\%} & 8.81\% \\
                           & Standard Deviation & 2.33 \% & \textbf{0.73\%} & 0.88\%  \\
  \hline         
  \end{tabular}
\end{table}

\subsection{Results with Transaction Costs}
\label{sec:Results}

The impact of transaction costs increases in the Black \& Scholes hedging strategy as the number of hedging points increases. This effect is apparent in both the average trading costs and the standard deviation of those costs. In contrast, the neural network, by optimizing the standard deviation of the hedging costs, can more effectively adapt the hedging frequency, thereby offering better control over both the mean and variability of those costs. With only 30 hedging points, the advantage of the neural network over the standard Black \& Scholes method becomes noticeable only when transaction costs reach 1\% or higher. However, with 90 hedging points, its benefits become apparent for costs of 0.5\% and above.
The Leland formula, arguably the best-known alternative to Black \& Scholes in the presence of transaction costs, exhibits similar characteristics. When the hedging frequency is as low as 30 points, the improvements offered by the Leland formula over Black \& Scholes are minimal for transaction costs below 1\%. With 90 hedging points, the Leland formula provides a clear improvement over Black \& Scholes for transaction costs of 0.5\% and above, and the neural network further improves the Leland approach for transaction costs at 0.5\% and above, despite the small number of observation points ($256 \times 90$) used to train it.

\section{Calibration on Even Smaller Datasets}
We now study how to calibrate the model using even fewer than 256 independent sequences by taking a single sequence of length 285 and constructing 256 overlapping sequences of length 30 from it. We first apply this procedure to simulated data and then to real-time data from the S\&P 500.

\subsection{Overlapping Sequences from Simulated Data}
\label{sec:Overlapping}
\begin{figure}[h]
    \centering
    \includegraphics[width=0.7\textwidth]{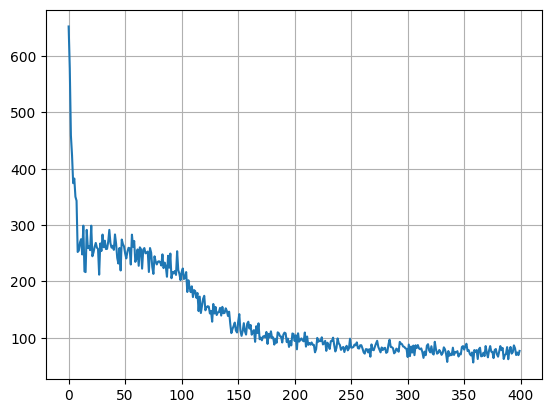}
    \caption{Calibration Overlapping Sequences, as discussed in section \ref{sec:Overlapping}: standard deviations in basis points of the hedging costs $Z_T$ on batches, with 30 hedging points and $T=0.25$ for the 500 epochs .}
    \label{fig:losscorrel}
\end{figure}

Here, a single sequence of length 285 is simulated, and 256 overlapping sequences of length 30 derived from it are used to train the neural network for hedging, with and without transaction costs.
The 256 resulting trajectories exhibit a minimum volatility of 14.29\%, a maximum volatility of 24.43\%, an average volatility of 19.43\%, and a standard deviation of 2.13\% for the realized volatilities.
The calibration speed of the neural network is comparable to that achieved when training on independent sequences of length 30, as shown in Figure~\ref{fig:losscorrel}.
Similarly, the hedging accuracy - evaluated on both the training and test sets, as presented in Tables~\ref{tab:3} and~\ref{tab:4} - is also comparable.

\begin{table}[h]
  \centering
  \caption{ Overlapping Sequences, as discussed in section \ref{sec:Overlapping}. 
  \\  Call Strike 100\%, $T=0.25$,  freq = 30. 
\\ Mean and Standard Deviation for $Z_T$ when $\sigma = 20\%$ and $r=0$.} 
  \vspace{5pt}  
  \label{tab:3}
  \begin{tabular}{|c|c|c|c|c|}
    \hline
    $\alpha$ & Costs for 256 Paths & Black \& Scholes & \textbf{Our Neural}     & Leland \\
        &  &  & \textbf{Network}     &  \\
    \hline
    \multicolumn{5}{|c|}{\textbf{Train Set}} \\
        \hline
    \multirow{2}{*}{0\% } & Mean & 3.99\% & 3.90\% & 3.99\% \\
                           & Standard Deviation & 0.69\% & 0.73\% & 0.69\%  \\
  \hline
          \multirow{2}{*}{2\% } & Mean & 7.76\% & \textbf{6.55\%} & 7.13\% \\
                           & Standard Deviation & 1.59\% & \textbf{0.74\%} & 0.97\%  \\
  \hline
       \multicolumn{5}{|c|}{\textbf{Test Set}} \\   
      \hline
    $\alpha$ & Costs for 256 Paths & Black \& Scholes & \textbf{Our Neural}     & Leland \\
        &  &  & \textbf{Network}     &  \\
    \hline
    \multirow{2}{*}{0\% } & Mean & 4.02\% & 3.97\% & 4.02\% \\
                           & Standard Deviation & 0.65\% & 0.88\% & 0.65\%  \\
  \hline
          \multirow{2}{*}{2\% } & Mean & 7.66\% & \textbf{6.80\%} & 7.14\% \\
                           & Standard Deviation & 1.61 \% & \textbf{1.01\%} & 1.00\%  \\
  \hline
  \end{tabular}
\end{table}

\begin{table}[ht]
  \centering
  \caption{ Overlapping Sequences as discussed is section \ref{sec:Overlapping}.
  \\ Call Strike 100\%, $T=0.25$,  freq = 90. 
    \\ Mean and Standard Deviation for $Z_T$ when $\sigma = 20\%$ and $r=0$.} 
  \vspace{5pt}  
  \label{tab:4}
  \begin{tabular}{|c|c|c|c|c|}
    \hline
    $\alpha$ & Costs for 256 Paths & Black \& Scholes & \textbf{Our Neural}     & Leland \\
        &  &  & \textbf{Network}     &  \\
    \hline
        \multicolumn{5}{|c|}{\textbf{Train Set}} \\
        \hline
    \multirow{2}{*}{0\% } & Mean & 3.85\% & \textbf{3.32\%} & 3.85\% \\
                           & Standard Deviation & 0.26\% & \textbf{0.25\%} & 0.26\%  \\
  \hline
          \multirow{2}{*}{2\% } & Mean & 10.52\% & \textbf{7.38\%} & 8.78\% \\
                           & Standard Deviation & 2.25 \% & \textbf{0.30\%} & 0.78\%  \\
  \hline
        \multicolumn{5}{|c|}{\textbf{Test Set}} \\
        \hline
         \multirow{2}{*}{0\% } & Mean & 4.02\% & 3.95\% & 4.02\% \\
                           & Standard Deviation & 0.38\% & 0.76\% & 0.38\% \\
  \hline
          \multirow{2}{*}{2\% } & Mean & 10.27\% & \textbf{7.98\%} & 8.87\% \\
                           & Standard Deviation & 2.43 \% & \textbf{0.88\%} & 0.99\%  \\
  \hline

  \end{tabular}
\end{table}

This result is, in our view, very important, as it suggests that neural networks - at least in their simplest form, as presented here - can be considered for use on real datasets, as a limited number of observations appears to be sufficient to train them.

\subsection{Overlapping Sequences from the S\&P 500}
Here, a single sequence of length 286 is constructed from the closing prices of the S\&P 500 Total Return Index over the period from 01/01/2024 to 25/02/2025. As before, 256 sequences of 31 data points are extracted from this dataset and used to train the neural network to price options with a maturity of $T=0.25$, assuming discrete hedging with 30 data points. Since the index is a total return index, there are no dividends to account for. We assume a constant instantaneous interest rate of 4.5\%, which is the average Secured Overnight Financing Rate (SOFR) in the US for the period. After calibration, the model produces unrealistic deltas that are not even close to the delta calculated using the Black \& Scholes model with the average 30-day volatility observed over the period. It seems that there is a need to introduce additional variables in the neural network to better handle situations with changing volatilities over time. We did not attempt to implement other models that could rely on forecasting volatility through other instruments, as we would have needed reliable quotes in such cases, but we leave this possibility open for future studies.

\section{Conclusion}
We show here that when asset prices follow a simple process such as geometric Brownian motion (GBM), as in the Black \& Scholes model, a neural network can be trained on a very small number of data points to find an optimal hedging strategy. This strategy can easily outperform a discretized Black \& Scholes model and Leland's model in the presence of transaction costs.

We also found that training can be effectively performed using overlapping sequences - an approach that is not often discussed in the literature. In our experiments, it was not necessary to simulate trajectories under a risk-neutral probability; instead, we simulated under a probability measure where the drift does not need to match the risk-free rate.

The fact that we only needed a limited amount of raw data for calibration - and that we did not need to renormalize the data - suggests that it may be possible to price some options using raw data alone, without relying on complex preprocessing or data enrichment, as mentioned in \citet{Mikkila} for 5-day options priced directly from intraday observations and featuring 7 hedging points a day.

That said, the neural network we use is very simple and does not include inputs such as market implied volatility (e.g., from volatility swaps) or realized volatility over a historical window. These could potentially help the model adapt to changing volatility regimes. As a result, the model did not produce satisfactory results when applied to trading S\&P 500 options using one year of daily historical data, largely due to regime changes in underlying volatility.

However, we expect that incorporating additional inputs into the neural network in future work will significantly improve its performance in these more complex cases.

\newpage

\appendix
\section{Appendix}
\label{appendix:proof}
We recall here for completeness Leland's formula \citep{Leland} for the impact of discretization on hedging, in the absence of transaction costs. From Equation (\ref{demo:appendix}), when hedging at observation points $t_i$ with $t_{i+1}-t_i = \Delta t = \frac{T}{n}$ we have,

\begin{equation}
\begin{aligned}
Z_T &= \sum\limits_{i=0}^{n-1}     e^{-rt_{i+1}}      \frac{\partial C_{\nu}}{\partial x}(X_{t_i},t_i) 
  \Big[X_{t_{i+1} } - e^{-r(t_{i+1}-t_i )}X_{t_i}  \Big] - e^{-rT}C_T 
\\&  = \sum\limits_{i=0}^{n-1}     e^{-rt_{i+1}}      \frac{\partial C_{\nu}}{\partial x}(X_{t_i},t_i) 
  \Big[X_{t_{i+1} } - e^{-r(t_{i+1}-t_i )}X_{t_i}  \Big] - \sum\limits_{i=0}^{n-1}  ( e^{-rt_{i+1}} C^{\nu}_{t_{i+1}}  - e^{-rt_{i}} C^{\nu}_{t_{i}} )  -  C^{\nu}_0
\end{aligned}
\end{equation}

We note 
$$ A_i= e^{-rt_{i+1}}      \frac{\partial C_{\nu}}{\partial x}(X_{t_i},t_i)  \Big(X_{t_{i+1} } - e^{-r(t_{i+1}-t_i )}X_{t_i}  \Big)$$ and 
$$B_i= e^{-rt_{i+1}}C^{\nu}_{t_{i+1}}- e^{-rt_i}C^{\nu}_{t_i}. $$

According to Equation (\ref{eq:C}) we have,
$$B_i=  \int_{t_i}^{t_{i+1}} e^{-rs}  \frac{\partial C_{\nu}}{\partial x}(X_s,s)(dX_s-rX_sds) \\
   + \frac{1}{2} \int_{t_i}^{t_{i+1}} e^{-rs} ( \sigma^2 - \nu^2  )X_s^2\frac{\partial^2 C_{\nu}}{\partial^2 x}(X_s,s)ds $$

$$=  \int_{t_i}^{t_{i+1}}  e^{-rt_{i+1}}  \frac{\partial C_{\nu}}{\partial x}(X_s,s)dX_s -  e^{-rt_{i+1}}  \frac{\partial C_{\nu}}{\partial x}(X_{t_t},t_i) rX_{t_i}\Delta t$$
$$  + \frac{1}{2}  e^{-rt_{i+1}} ( \sigma^2 - \nu^2  )X_{t_i}^2\frac{\partial^2 C_{\nu}}{\partial^2 x}(X_{t_i},t_i) \Delta t +O(\Delta t^{\frac{3}{2}})$$ 

On the other hand,
$$A_i=e^{-rt_{i+1}} \frac{\partial C_{\nu}}{\partial x}(X_{t_i},t_i)  (X_{t_{i+1} }-X_{t_i} - r X_{t_i}\Delta t  ) +O(\Delta t^2)$$
So,
$$A_i-B_i = e^{-rt_{i+1}}   \int_{t_i}^{t_{i+1}} \Big(   \frac{\partial C_{\nu}}{\partial x}(X_{t_i},t_i)  -\frac{\partial C_{\nu}}{\partial x}(X_s,s)       \Big) dX_s $$ 
$$  + \frac{1}{2}  e^{-rt_{i+1}} ( \nu^2 -\sigma^2  )X_{t_i}^2\frac{\partial^2 C_{\nu}}{\partial^2 x}(X_{t_i},t_i) \Delta t +O(\Delta t^{\frac{3}{2}})$$

Now, 
\begin{equation}  
\label{eq:derivative}
\frac{\partial C_{\nu}}{\partial x}(X_{t_i},t_i)
-  \frac{\partial C_{\nu}}{\partial x}(X_s, s) 
=    \frac{\partial^2 C_{\nu}}{\partial^2 x} (X_{t_i},t_i) (X_{t_i}-X_s) +O(\Delta t)
\end{equation}  

and
$$ \int_{t_i}^{t_{i+1}}   (X_{t_i}-X_{t_s})  dX_s   =  - \frac{1}{2} (X_{t_{i+1}} - X_{t_i})^2 +   \frac{1}{2}  \int_{t_i}^{t_{i+1}}  \sigma^2 X_s^2 ds  $$
$$ = - \frac{1}{2} \Big[(X_{t_{i+1}} - X_{t_i})^2 -\sigma^2 X_{t_i}^2 \Delta t\Big] +O(\Delta t ^{\frac{3}{2}})$$
So, 

$$A_i -B_i =    \frac{1}{2}  e^{-rt_{i+1}}  \frac{\partial^2 C_{\nu}}{\partial^2 x} (X_{t_i},t_i) X_{t_i}^2 \Big[   \sigma^2 \Delta t  - (X_{t_{i+1}} - X_{t_i})^2        \Big]           $$
$$  + \frac{1}{2}  e^{-rt_{i+1}} ( \nu^2 -\sigma^2  )X_{t_i}^2\frac{\partial^2 C_{\nu}}{\partial^2 x}(X_{t_i},t_i) \Delta t +O(\Delta t^{\frac{3}{2}})$$
 and finally, 
\begin{equation} \label{eq:approx}
 Z_T=    \frac{1}{2}  \sum\limits_{i=1}^n e^{-rt_{i+1}} \frac{\partial^2 C_{\nu}}{\partial^2 x} (X_{t_i},t_i) X_{t_i}^2  \left[  \nu^2 \Delta t -\left(\frac{X_{t_{i+1}} - X_{t_i}}{X_{t_i}}\right)^2  \right]  - C_0^{\nu} +O(\sqrt{\Delta t})
 \end{equation}
 which is equivalent to Leland's Formula \citep{Leland}.

\section{Appendix}
\label{appendix:proof2}

We consider the log normal model:
\begin{align}
 dX_t = \mu X_tdt + \sigma X_t dW_t  \mbox{ with } \sigma>0,
\end{align}

where $(W_t)_{t\geq 0}$ is a Brownian motion in the probability space $(\Omega, P)$.\\
We provide here a short proof of the result by \citet{Soner} which states that for a fixed implied volatility $\nu$ (not dependent on $\Delta t$), if one uses the Black \& Scholes deltas,  the transactions costs incurred tend to infinity when $\Delta t$ tends to zero. We demonstrate the limit almost surely. Without loss of generality, we consider an option of maturity $T=1$. 
We note 

\begin{equation}f(s,b)= exp\left(\sigma b +(\mu - \frac{\sigma^2}{2})s\right),\end{equation}
and we have, 
$$X_s = f(s,W_s).$$

For $x \in \R$ we note $[x]$ the integer part of $x$.

\begin{lemma} $\forall T >0$
\begin{equation} \label{eq1:lemma}
\sum\limits_{i=0}^{[nT]-1} \left|W_{\frac{i+1}{n}}-W_{\frac{i}{n}}\right|^2 \xrightarrow[n \to \infty]{} T \quad  \mbox{ almost surely.} \end{equation}
\begin{equation} \label{eq2:lemma} \sum\limits_{i=0}^{[nT]-1} \left|W_{\frac{i+1}{n}}-W_{\frac{i}{n}}\right| \xrightarrow[n \to \infty]{} + \infty \mbox{ almost surely.} \end{equation} 
\label{lemma}
\end{lemma}

\begin{proof}

The maximum lengths of the intervals in step $n$ that form the partition $0<\frac{1}{n}< \cdots < \frac{[nT]}{n}$  are $\lambda_n =\frac{1}{n}$ and satisfy the condition
$\lambda_n =o (1/log(n))$, which according to \cite[Thm. 4.5 page 89]{Dudley} 
proves the almost sure convergence in Equation (\ref{eq1:lemma}).

\vspace{0.5 cm}
For the second  point, almost surely:
\begin{equation} 
\label{qv}
\sum\limits_{i=0}^{[nT]-1} \left|W_{\frac{i+1}{n}}-W_{\frac{i}{n}}\right|^2 \leq
\sup\limits_{|s-t|<\frac{1}{n}, s,t \in [0,T]} \left|W_s-W_t\right|\cdot
\sum\limits_{i=0}^{[nT]-1} \left|W_{\frac{i+1}{n}}-W_{\frac{i}{n}}\right|.
\end{equation}

Now, for any $\omega \in \Omega$, $W_s(\omega)$ is continuous in $s$ in $[0,T]$, which is compact, therefore it is also absolutely continuous and,
$$\sup\limits_{|s-t|<\frac{1}{n}, s,t \in [0,T]} \left|W_s(\omega)-W_t(\omega)\right| 
\xrightarrow[n \to \infty]{}  0 \mbox{ almost surely,} $$
and as 
$$\sum\limits_{i=0}^{[nT]-1} \left|W_{\frac{i+1}{n}}-W_{\frac{i}{n}}\right|^2 
\xrightarrow[n \to \infty]{}  T \mbox{ almost surely,} $$

then necessarily, 
$\sum\limits_{i=0}^{[nT]-1} \left|W_{\frac{i+1}{n}}-W_{\frac{i}{n}}\right|$ tends to $+ \infty$, for the inequality of  Equation (\ref{qv}) to be satisfied. Therefore, Equation (\ref{eq2:lemma}) is satisfied.

\end{proof}

The transaction costs are $\alpha V_n$ with $\alpha>0$ and,
$$V_n = \sum\limits_{i=0}^{n-1}  \left| \frac{\partial C}{\partial x}\left(X_{\frac{i+1}{n}} , \frac{i+1}{n}\right)- \frac{\partial C}{\partial x}\left(X_{\frac{i}{n}}, \frac{i}{n}\right)\right|X_{\frac{i+1}{n}}.$$ According to Taylor's formula with integral residue:

$$ \frac{\partial C}{\partial x}(X_{\frac{i+1}{n}}, \frac{i+1}{n})- \frac{\partial C}{\partial x}(X_{\frac{i}{n}}, \frac{i}{n})$$
$$ = \frac{\partial^2 C}{\partial^2 x}(X_{\frac{i}{n}}, \frac{i}{n})\left(X_{\frac{i+1}{n}}-X_{\frac{i}{n}}\right)
+ \frac{\partial^2 C}{\partial x \partial t}(X_{\frac{i}{n}}, \frac{i}{n})\frac{1}{n}$$
$$+ \int_0^1 (1-s) \Delta_{i,n}^{T} H\left(\frac{\partial C}{\partial x}\right) 
\left(X_{\frac{i}{n}}+ s(X_{\frac{i+1}{n}}-X_{\frac{i}{n}}), \frac{i+s}{n} \right) \Delta_{i,n}ds.$$
where  $ \Delta_{i,n}^T = (X_{\frac{i+1}{n}}-X_{\frac{i}{n}} , \frac{1}{n})  $ and $H$ is the Jacobian operator.

\vspace{0.2 cm}
Let $D_{\eta}= \{(x,s) , |x-X_0|<\eta \mbox{ and } |s|<\eta \}$
and let $\epsilon >0$  be such that $\forall (x,s) \in D_{\epsilon}$ we have:
\[
\left\{
\begin{array}{ll}
\frac{\partial^2 C}{\partial^2 x}(x,s) \geq  
\frac{1}{2} \frac{\partial^2 C}{\partial^2 x}(X_0,0)\\ \\
|\frac{\partial^2 C}{\partial x \partial t}(x,s)| \leq 1+ |\frac{\partial^2 C}{\partial x \partial t}(X_0,0)|\\ \\
\|H(\frac{\partial C}{\partial x})(x,s)\| \leq 1+ \|H(\frac{\partial C}{\partial x})(X_0,0) \|, 
\end{array}
\right.
\]
where $\| \cdot\|$ is the spectral norm for which $|z^{T}Mz| \leq \| M\| \|z\|^2$ (here $z^T$ is the transpose of $z$); note that $\epsilon$ exists because of the continuity of the functions considered. Note also that $D_{\epsilon }$ is convex and therefore,
$$(X_{i/n},i/n) \mbox{ and } (X_{(i+1)/n},(i+1)/n)  \in D_{\epsilon}$$
$$ \Longrightarrow \forall s \in [0,1], (X_{i/n}+s ( X_{(i+1)/n}- X_{i/n}   ), (i+s)/n ) \in D_{\epsilon}.$$

\vspace{0.2 cm}
Now, let 
$\tau_1= \inf \{s\in \R^+ , (X_s,s) \notin D_{\epsilon} \mbox{ or } X_s \notin [\frac{1}{2} |X_0| ,2|X_0| ]\}$ 
then, 
$$V_n \geq \sum\limits_{i=0}^{n-1} 1_{\frac{i+1}{n} 
\leq \tau_1} \left| \frac{\partial C}{\partial x}(X_{\frac{i+1}{n}, \frac{i+1}{n}}) - \frac{\partial C}{\partial x}(X_{\frac{i}{n}, \frac{i}{n}})\right| X_{\frac{i+1}{n}}$$

$$ \geq A_n + B_n +C_n,$$

where,
\begin{align} \label{eq:An}
& A_n = \sum\limits_{i=0}^{n-1} 1_{\frac{i+1}{n}\leq \tau_1} \frac{X_0}{4} \cdot \frac{\partial^2 C}{\partial^2 x} (X_0, 0) \cdot \left| X_{\frac{i+1}{n}} - X_{\frac{i}{n}} \right|, \\
& B_n = - \sum\limits_{i=0}^{n-1} 1_{\frac{i+1}{n}\leq \tau_1} 
 \frac{2X_0}{n} \cdot \left(1+ \left| \frac{\partial^2 C}{\partial x \partial t}(X_0, 0) \right| \right),  
\\ & C_n = - \sum\limits_{i=0}^{n-1} 1_{\frac{i+1}{n}\leq \tau_1} 
 2X_0 \cdot \left(1+\left\| H\left(\frac{\partial C}{\partial x}\right)(X_0,0) \right\| \right)
\left( \int\limits_0^1 (1-s) ds \right)
\left(\frac{1}{n^2} +  \left| X_{\frac{i+1}{n}} - X_{\frac{i}{n}} \right|^2\right).
\end{align}

For $B_n$ we have:
\begin{equation} B_n \geq -2 X_0 \cdot \left(1+\left|\frac{\partial^2 C}{\partial x \partial t}(X_0,0)\right|\right).
\end{equation}

For $C_n$ we have: 
$$- \sum\limits_{i=0}^{n-1} 1_{\frac{i+1}{n}\leq \tau_1} \cdot \frac{1}{n^2} \geq - \frac{1}{n},$$
and

$$
- \sum\limits_{i=0}^{n-1} 1_{\frac{i+1}{n}\leq \tau_1} 
\left|X_{\frac{i+1}{n}}-X_{\frac{i}{n}}\right|^2  
\geq  - \sum\limits_{i=0}^{n-1}  \left|X_{\frac{i+1}{n}}-X_{\frac{i}{n}}\right|^2. 
$$
Let, 
\begin{equation}
\label{eq:sum}
S_n=  - \sum\limits_{i=0}^{n-1} 
\left|X_{\frac{i+1}{n}}-X_{\frac{i}{n}}\right|^2. 
\end{equation}

According to the mean value theorem,
$\exists$ $\xi_{i,n} \in \left[ \left( \frac{i}{n} , W_{\frac{i}{n}}\right),\left( \frac{i+1}{n}, W_{\frac{i+1}{n}}\right)\right]:$
\begin{equation} \label{MVT}
X_{\frac{i+1}{n}}-X_{\frac{i}{n}} = \nabla f(\xi_{i,n}). 
\begin{pmatrix} 
\frac{1}{n} \\ 
W_{\frac{i+1}{n}} -W_{\frac{i}{n} } 
\end{pmatrix},
\end{equation}
where $\left[ \left( \frac{i}{n} , W_{\frac{i}{n}}\right),\left( \frac{i+1}{n}, W_{\frac{i+1}{n}}\right)\right]$ is the $\R^2$ the segment between the two points 
$\left( \frac{i}{n} , W_{\frac{i}{n}}\right) $ and $\left( \frac{i+1}{n} , W_{\frac{i+1}{n}}\right) $.

\vspace{0.5cm}
Let $\omega \in \Omega$ and  
$\mathcal{B}(w)= [0,1] \times [\inf\limits_{s \in [0,1]}  W_s(w) , \sup\limits_{s \in [0,1]}W_s(w)] $ 
and 
\begin{equation} M(w) = \sup\limits_{\xi \in \mathcal{B}(w) } \| \nabla f(\xi) \|.
\end{equation}
The $\sup$ is reached as $\mathcal{B}(w)$  is compact and $\nabla f$ is continuous. Therefore, 
$$M(\omega) < +\infty.$$ Using Cauchy Schwartz, 
$$
\forall n \in \mathbb{N},\ \forall i \in \llbracket 0,\ n - 1 \rrbracket, 
|    X_{\frac{i+1}{n}} -X_{\frac{i}{n}}    |^2 \leq M^2 
\left( \frac{1}{n^2} + |W_{\frac{i+1}{n}} -W_{\frac{i}{n}}    |^2   \right) 
$$

So,
$$S_n \geq  - \frac{M^2}{n } - M^2 \sum\limits_{i=0}^{n-1} 
\left|W_{\frac{i+1}{n}}-W_{\frac{i}{n}}\right|^2,$$

and therefore according to Lemma \ref{lemma}.
$$
\lim\inf\limits_{n \longrightarrow  +\infty} S_n \geq -M^2 T \mbox{ almost surely},
$$
and as a consequence, 
\begin{equation}
\lim\inf\limits_{n \longrightarrow  +\infty} C_n \geq -X_0 \left(1+\left\| H\left(\frac{\partial C}{\partial x}\right)(X_0,0) \right\| \right)M^2 T \mbox{ almost surely}.
\end{equation}

\vspace{0.5cm}

For $A_n$ defined in Equation (\ref{eq:An}), 
let $\tilde{D}_{\eta}= \{(s,w) , |w|<\eta \mbox{ and } |s|<\eta \}$ and 
let $\tilde{\epsilon}>0$ be such that $\forall \xi \in  \tilde{D}_{\tilde{\epsilon}}$ 
\begin{equation} \label{ineq2}
\left\{
\begin{array}{ll}
\frac{\partial f}{\partial b}(\xi) \geq  
\frac{1}{2}\frac{\partial f}{\partial b}(0) = \frac{\sigma}{2}\\ \\
|\frac{\partial f}{\partial s}(\xi)| \leq 1+ |\frac{\partial f}{\partial s}(0)|. 
\end{array}
\right.
\end{equation}

Now, let 
$\tau= \inf \{s\in \R^+ , (s, W_s) \notin \tilde{D}_{\tilde{\epsilon}}\}$, let $\tau_2 = \inf(\tau_1,\tau)$ and let 
$$\Theta_m = \left\{ \omega \in \Omega, \tau_2 \geq  \frac{1}{m} \right\}.$$

We have $\bigcup\limits_{m \in \mathbb{N}} \Theta_m = \Omega $ almost surely by the continuity of $X_s$ and $W_s$.
Let, 
\begin{equation}V_n= \sum\limits_{i=0}^{n-1} 1_{\frac{i+1}{n}\leq \tau_1} \left|X_{\frac{i+1}{n}}-X_{\frac{i}{n}}\right|,
\end{equation}
then, $ \forall m\in \mathbb{N}, \forall n\in \llbracket m, +\infty \rrbracket$,
$$ V_n 
\geq   1_{\tau_2\geq \frac{1}{m}} \sum\limits_{i=0}^{n-1} 1_{\frac{i+1}{n}\leq \tau_1}  \left|X_{\frac{i+1}{n}}-X_{\frac{i}{n}}\right|
 \geq  \sum\limits_{i=0}^{n-1} 
 1_{\frac{i+1}{n}\leq \frac{1}{m}} 1_{\tau_2\geq \frac{1}{m}} \left|X_{\frac{i+1}{n}}-X_{\frac{i}{n}}\right|.
$$

Now, according to Equation (\ref{MVT}) and Equation (\ref{ineq2}), 
$$  1_{\frac{i+1}{n}\leq \frac{1}{m}\leq \tau_2}\left|X_{\frac{i+1}{n}}-X_{\frac{i}{n}}\right|  \geq 
 1_{\frac{i+1}{n}\leq \frac{1}{m} \leq \tau_2} 
\left(  \frac{\sigma}{2} \left|W_{\frac{i+1}{n}}-W_{\frac{i}{n}}\right|       
- \left(1+ |\frac{\partial f}{\partial s}(0)| \right) \frac{1}{n} \right), $$

So,
\begin{equation} 
V_n \geq 
- \left(1+ |\frac{\partial f}{\partial s}(0)| \right) +
1_{\tau_2\geq \frac{1}{m}} 
\frac{\sigma}{2}
\sum\limits_{i=0}^{n-1} 1_{\frac{i+1}{n}\leq \frac{1}{m}} \left|W_{\frac{i+1}{n}}-W_{\frac{i}{n}}\right|. 
\end{equation}

Now,
\begin{equation}
\sum\limits_{i=0}^{n-1} 1_{\frac{i+1}{n}\leq \frac{1}{m}} \left|W_{\frac{i+1}{n}}-W_{\frac{i}{n}}\right|   
= \sum\limits_{i=0}^{[\frac{n}{m}]-1}  \left|W_{\frac{i+1}{n}}-W_{\frac{i}{n}}\right|, 
\end{equation}
and according to Lemma (\ref{lemma}): 
$$\forall m \in \mathbb{N}, \sum\limits_{i=0}^{[\frac{n}{m}]-1}  \left|W_{\frac{i+1}{n}}-W_{\frac{i}{n}}\right|   \xrightarrow[n \to \infty]{}  +\infty \mbox{ almost surely }. $$

So, 
\begin{equation}
\label{eq:final}
\forall \omega \in \Omega, \forall m\in \mathbb{N}: 
\lim\limits_{n \longrightarrow +\infty}V_n \geq  
\left( 1_{\tau_1\geq \frac{1}{m}} \right) 
\cdot (+ \infty),
\end{equation}

but: 
$$\lim\limits_{m \longrightarrow +\infty}   1_{\tau_1\geq \frac{1}{m}}  =1_{\Omega}
\mbox{ almost surely}.$$

So, taking the limit in $m$ in Equation (\ref{eq:final}), we get

$$\lim\limits_{n \longrightarrow +\infty} V_n = +\infty
\mbox{ almost surely},$$

and therefore, 
$$\lim\limits_{n \longrightarrow +\infty} A_n = +\infty
\mbox{ almost surely},$$

which finishes the proof.


\end{document}